\documentclass[aps,prb,times,twocolumn,amsmath,amssymb,superscriptaddress,floatfix,showpacs]{revtex4}

\usepackage{color}

\usepackage{graphicx}
\usepackage{subfigure}
\usepackage{bbold}
\usepackage{verbatim}
\usepackage{float}
\usepackage{enumerate}
\usepackage{amsfonts}
\usepackage{multirow}
\usepackage[colorlinks,bookmarks=false,citecolor=blue,linkcolor=red,urlcolor=blue]{hyperref}

\begin{document}


\title{
Disorder driven spin-orbital liquid behaviour in the Ba$_3$XSb$_2$O$_9$ materials

}


\author{Andrew Smerald}

\affiliation{Institut de Th{\'e}orie des Ph{\'e}nom\`{e}nes Physiques, Ecole Polytechnique F{\'e}d{\'e}rale de Lausanne (EPFL), CH-1015 Lausanne, Switzerland}

\author{Fr{\'e}d{\'e}ric Mila}
\affiliation{Institut de Th{\'e}orie des Ph{\'e}nom\`{e}nes Physiques, Ecole Polytechnique F{\'e}d{\'e}rale de Lausanne (EPFL), CH-1015 Lausanne, Switzerland}


\date{\today}


\begin{abstract}  
Recent experiments on the Ba$_3$XSb$_2$O$_9$ family have revealed materials that potentially realise spin- and spin-orbital liquid physics.
However, the lattice structure of these materials is complicated due to the presence of charged X$^{2+}$-Sb$^{5+}$ dumbbells, with two possible orientations.
To model the lattice structure, we consider a frustrated model of charged dumbbells on the triangular lattice, with long-range Coulomb interactions.
We study this model using Monte Carlo simulation, and find a freezing temperature, $T_{\sf frz}$, at which the simulated structure factor matches well to low-temperature x-ray diffraction data for Ba$_3$CuSb$_2$O$_9$.
At $T=T_{\sf frz}$ we find a complicated ``branching'' structure of superexchange-linked X$^{2+}$ clusters, which form a fractal pattern with fractal dimension $d_{\sf f}=1.90$.
We show that this gives a natural explanation for the presence of orphan spins.
Finally we provide a plausible mechanism by which such dumbbell disorder can promote a spin-orbital resonant state with delocalised orphan spins.
\end{abstract}

\pacs{
75.10.Kt,	
75.25.Dk,	
75.47.Lx	
}

\maketitle


Recently there has been an intense search for materials exhibiting spin-liquid behaviour -- materials beyond the ``standard model'' of condensed matter physics\cite{lacroix}.
A particularly intriguing idea is of a spin-orbital liquid, in which not only the spin but also the orbital degrees of freedom remain fluctuating down to low temperature\cite{lacroix,ishihara97,khaliullin00,khomskii03,corboz12}.


The Ba$_3$XSb$_2$O$_9$ family, with X=Cu\cite{zhou11,nakatsuji12,quilliam12,ishiguro13,nasu13,katayama15,smerald14,shanavas14,do14,nasu15,han15},Ni\cite{cheng11,xu12,bieri12,chen12,hwang13},Co\cite{shirata12,susuki13,koutroulakis15},Mn\cite{tian14a,tian14b}\dots, has been shown to be a promising class of materials to realise spin-liquid behaviour.
Ba$_3$CuSb$_2$O$_9$ has been particularly well studied, and it has been suggested that the spin and orbital degrees of freedom associated with the Cu$^{2+}$ ions form a spin-orbital liquid state\cite{zhou11,nakatsuji12,quilliam12,ishiguro13,nasu13,katayama15,smerald14,shanavas14,do14,nasu15,han15}.
In the case of Ba$_3$NiSb$_2$O$_9$, the pressure-synthesised 6H-B structure has been proposed as an example of a spin-1 spin-liquid state\cite{cheng11,xu12,bieri12,chen12,hwang13}.


An important starting point when trying to understand spin-liquid behaviour is knowledge of the lattice structure.
In Ba$_3$CuSb$_2$O$_9$ it has been suggested that the Cu$^{2+}$ ions form a short-range honeycomb lattice\cite{nakatsuji12}, and theoretical approaches have therefore concentrated on Cu$^{2+}$ plaquettes formed of several hexagons\cite{nasu13,smerald14,nasu15}.
On the other hand, in the 6H-B phase of Ba$_3$NiSb$_2$O$_9$ it has been suggested that the Ni$^{2+}$ ions form a triangular lattice\cite{cheng11}.

\begin{figure}[t]
\centering
\includegraphics[width=0.49\textwidth]{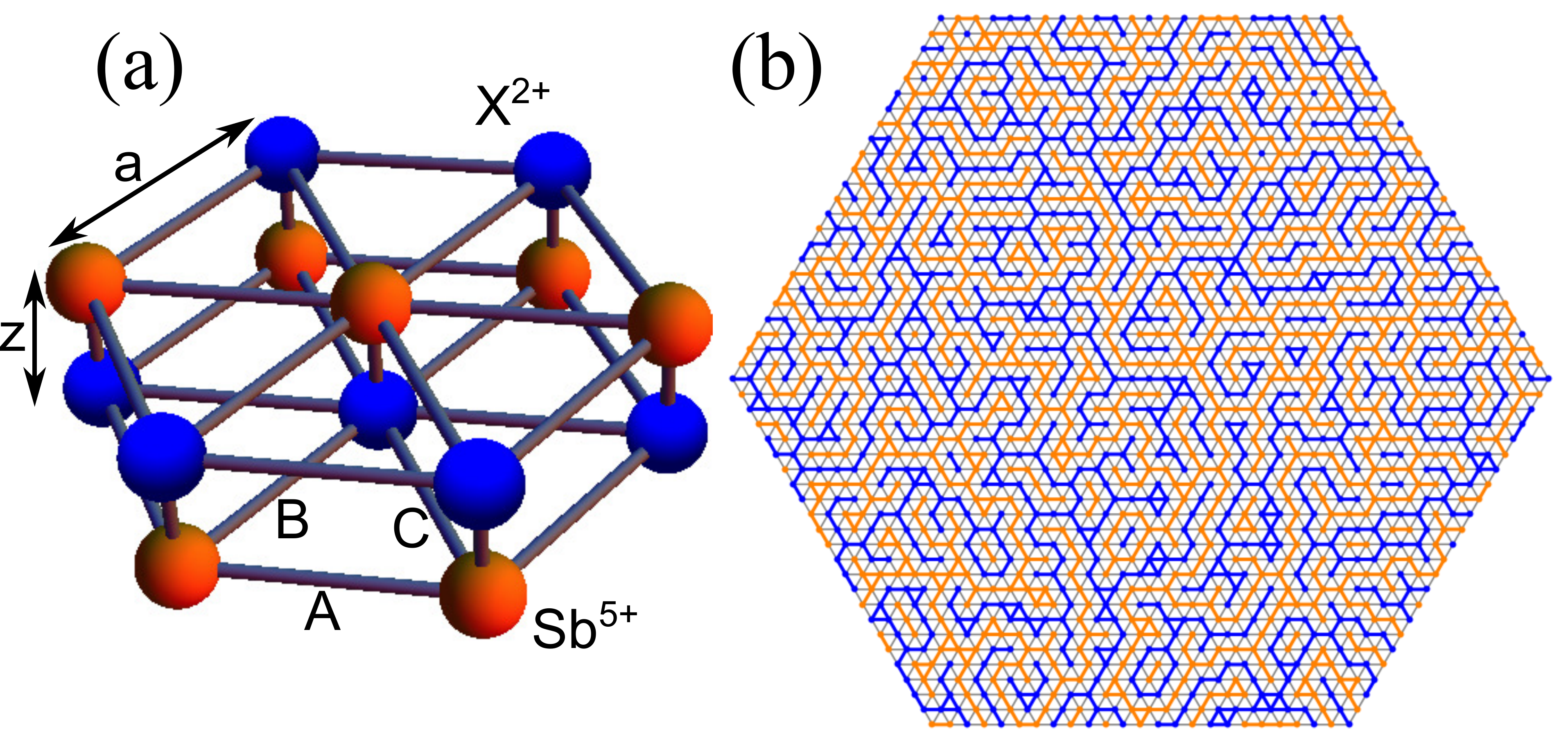}
\caption{\footnotesize{
Charged dumbbells on the triangular lattice. 
(a) X$^{2+}$-Sb$^{5+}$ dumbbells of length $z$ form a triangular lattice bilayer.
%
There is an Ising degree of freedom associated to whether the dumbbell is orientated with X above Sb or vice versa.
The equilibrium distribution of dumbbells can be mapped onto a charge model, $E_{\sf Coul} $ [Eq.~\ref{eq:CoulEn}], which at low temperature orders in a stripe ground state [shown here].
%
(b) Material realisations of $E_{\sf Coul} $ [Eq.~\ref{eq:CoulEn}] fall out of equilibrium at $T=T_{\sf frz}$, and the lattice structure can be studied by making simulations at this temperature.
A snapshot of a typical lattice structure for X=Cu is shown, with blue and orange sites denoting different dumbbell orientations.
Superexchange interactions link Cu$^{2+}$ ions on dumbbells with the same orientation, and superexchange linked clusters are shown by blue and orange bonds.  
}}
\label{fig:lattice}
\end{figure}

Here we argue that in neither case is this a good starting point for theoretical investigation, and instead one should consider a disordered ``branch'' lattice [see Fig.~\ref{fig:lattice}b].
The evidence we present focuses in particular on Ba$_3$CuSb$_2$O$_9$, but should be applicable to other members of the Ba$_3$XSb$_2$O$_9$ family. 
Furthermore, we suggest that this type of correlated lattice disorder can promote spin-orbital liquid behaviour.

%


In order to investigate the lattice structure of these materials, we solve a frustrated model of interacting X$^{2+}$-Sb$^{5+}$ charged dumbbells [see Fig.~\ref{fig:lattice}].
We argue this is relevant to stoichiometric X=Cu, X=Ni in the pressure synthesised 6H-B phase and potentially to pressure synthesised X=Mn and X=Co.
%

%
The X$^{2+}$-Sb$^{5+}$ dumbbells are surrounded by O$^{2-}$ bioctahedra, and their constituent ions sit on the vertices of stacked triangular lattice bilayers\cite{nakatsuji12}, as shown in Fig.~\ref{fig:lattice}a.
Each dumbbell has two possible orientations with either the X$^{2+}$ or Sb$^{5+}$ on top.
Electrostatically, the primary influence on the orientation of the dumbbells is the orientation of the other dumbbells -- that is to say that the Ba$^{2+}$, O$^{2-}$ and remaining Sb$^{5+}$ ions are electrostatically ambivalent as to the dumbbell orientation.

This leads us to consider a Coulombic charge model,
\begin{align}
E_{\sf Coul} =  \frac{1}{2} \sum_{i\neq j} \frac{q_iq_j}{r_{ij}},
\label{eq:CoulEn}
\end{align}
where $q_i =\pm 1$ is a normalised charge, $i$ and $j$ run over the sites of a bilayer triangular lattice [shown in Fig.~\ref{fig:lattice}], $r_{ij} = |{\bf r}_i - {\bf r}_j |$ and the charge distribution is constrained to have one positive and one negative charge on each dumbbell.
We ignore interaction between dumbbells in different bilayers, and we provide a justification for this approximation below.


In order to relate the charge distribution following from $E_{\sf Coul} $ [Eq.~\ref{eq:CoulEn}] to the lattice structure of the materials, it is necessary to understand the synthesis process.
This is typically performed at high temperature ($>$1000$^\circ$C), and the crystals are then slowly cooled to room temperature and below\cite{nakatsuji12,quilliam12}.
A characteristic timescale $t_{\sf cool}$ can be ascribed to this cooling process, and this should be compared to $t_{\sf flip}$, the characteristic time for dumbbells to reverse their orientation.
Close to the synthesis temperature, we assume that $t_{\sf flip}\ll t_{\sf cool}$, and therefore the dumbbell orientation remains in thermal equilibrium as $T$ is reduced.
As the crystal is cooled, $t_{\sf flip}$ increases, and there is a temperature, $T_{\sf frz}$, below which $t_{\sf flip} \gg t_{\sf cool}$.
In this regime the dumbbell dynamics is too slow to equilibriate the system and the charge distribution is thus frozen in place.
The dumbbell structure for any $T<T_{\sf frz}$ can therefore be understood from studying the equilibrium dumbbell structure at $T=T_{\sf frz}$.

The dumbbells in these materials are widely spaced, and one piece of evidence that they are dynamic at high temperature comes from the isostructural compound Ba$_3$IrTi$_2$O$_9$\cite{dey12}.
Here the Ir-Ti dumbbells exhibit a markedly different low-temperature structure depending on whether the material is slowly cooled from the synthesis temperature (1000$^\circ$C) or quenched. 
%

This suggests a twofold strategy for understanding the lattice structure of these materials.
1) Simulate $E_{\sf Coul} $ [Eq.~\ref{eq:CoulEn}] as a function of temperature, and, by comparison with experimental data, determine the freezing temperature, $T_{\sf frz}$.
2) Simulate the model at $T_{\sf frz}$ in order to extract detailed information about the lattice structure for all $T<T_{\sf frz}$.


In order to simulate $E_{\sf Coul} $ [Eq.~\ref{eq:CoulEn}], it is first mapped onto an Ising model on the triangular lattice using Ewald summation\cite{grzybowski00}.
This leads to,
\begin{align}
E_{\sf Coul} = E_0 + \frac{1}{2} \sum_{i,j} \psi_{ij}(z) \sigma_i \sigma_j,
\label{eq:EIs}
\end{align}
where $\sigma_i =\pm 1$ is an Ising spin, $i$ runs over the sites of a triangular lattice and $\psi_{ij}(z)$ defines the interactions between sites as a function of the dumbbell size, $z$ [see Fig.~\ref{fig:lattice} and Appendix~\ref{sec:psimatrix}].
%
%
%
For $z\to 0$, $E_{\sf Coul} $ [Eq.~\ref{eq:EIs}]  reduces to interacting Ising dipoles on the triangular lattice\cite{rossler01}.
Here we consider $z=0.46a$ as this is relevant to Ba$_3$CuSb$_2$O$_9$\cite{nakatsuji12}.

\begin{figure}[t]
\centering
\includegraphics[width=0.49\textwidth]{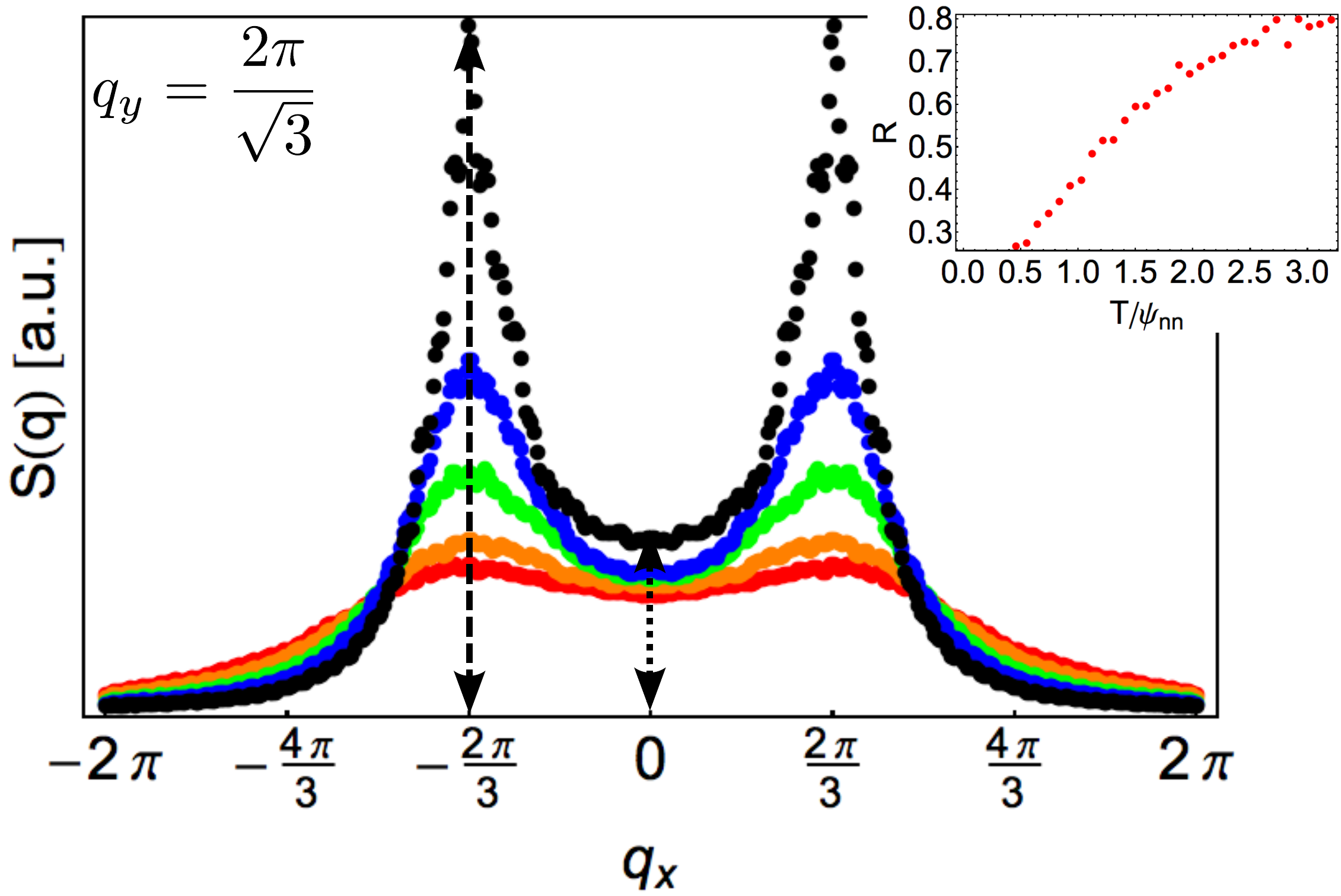}
\caption{\footnotesize{
The dumbbell structure factor, $S({\bf q})$, as predicted by simulations of $E_{\sf Coul} $ [Eq.~\ref{eq:CoulEn}].
The structure factor is plotted at a range of temperatures as a function of $q_{x}$ with $q_{y}=2\pi/\sqrt{3}$ and $q_{\sf z}=\pi$. 
From top to bottom: $T/\psi_{\sf nn}=0.45$ (black), $T/\psi_{\sf nn}=0.9$ (blue), $T/\psi_{\sf nn}=1.4$ (green), $T/\psi_{\sf nn}=2.4$ (orange) and $T/\psi_{\sf nn}=3.2$ (red). 
In the inset, the ratio \mbox{$R=S(0,2\pi/\sqrt{3},q_{\sf z})/S(2\pi/3,2\pi/\sqrt{3},q_{\sf z})$} (small dotted arrow compared to large dashed arrow) is plotted as a function of $T$. 
}}
\label{fig:ratio}
\end{figure}
%


We have performed Monte Carlo simulations of $E_{\sf Coul} $ [Eq.~\ref{eq:EIs}] over a wide range of temperatures [see Appendix~\ref{sec:MCdetails}].
%
%
The ground state is 6-fold degenerate, and consists of alternating stripes of $\sigma=1$ and $\sigma=-1$, parallel to either the {\sf A}, {\sf B} or {\sf C} bonds [see Fig~\ref{fig:lattice}a].
\footnote{
In a 3d model of charged dumbbells we expect the ground state to retain a stripe pattern within the layers.
For nearest-neighbour interlayer coupling, the stripe state gives the best possible interlayer energy consistent with the absence of defect triangles in the plane.
Other simple 2d states, such as the honeycomb, give much worse interlayer energies.
Thus we expect that the 3d model shows the same qualitative features as the 2d model.
}
At $T_{\sf c}/\psi_{\sf nn} \approx 0.19$, with the nearest-neighbour interaction $\psi_{\sf nn} \approx 0.18$ in the units of Eq.~\ref{eq:CoulEn}, there is an apparently 1st order phase transition into a domain wall network state, as proposed in Ref.~[\onlinecite{korshunov05}] for the Ising model with further neighbour exchange interactions.
We postpone a detailed description of the low-temperature behaviour to another publication, and instead concentrate on the temperature region above the phase transition.

\begin{figure}[t]
\centering
\includegraphics[width=0.49\textwidth]{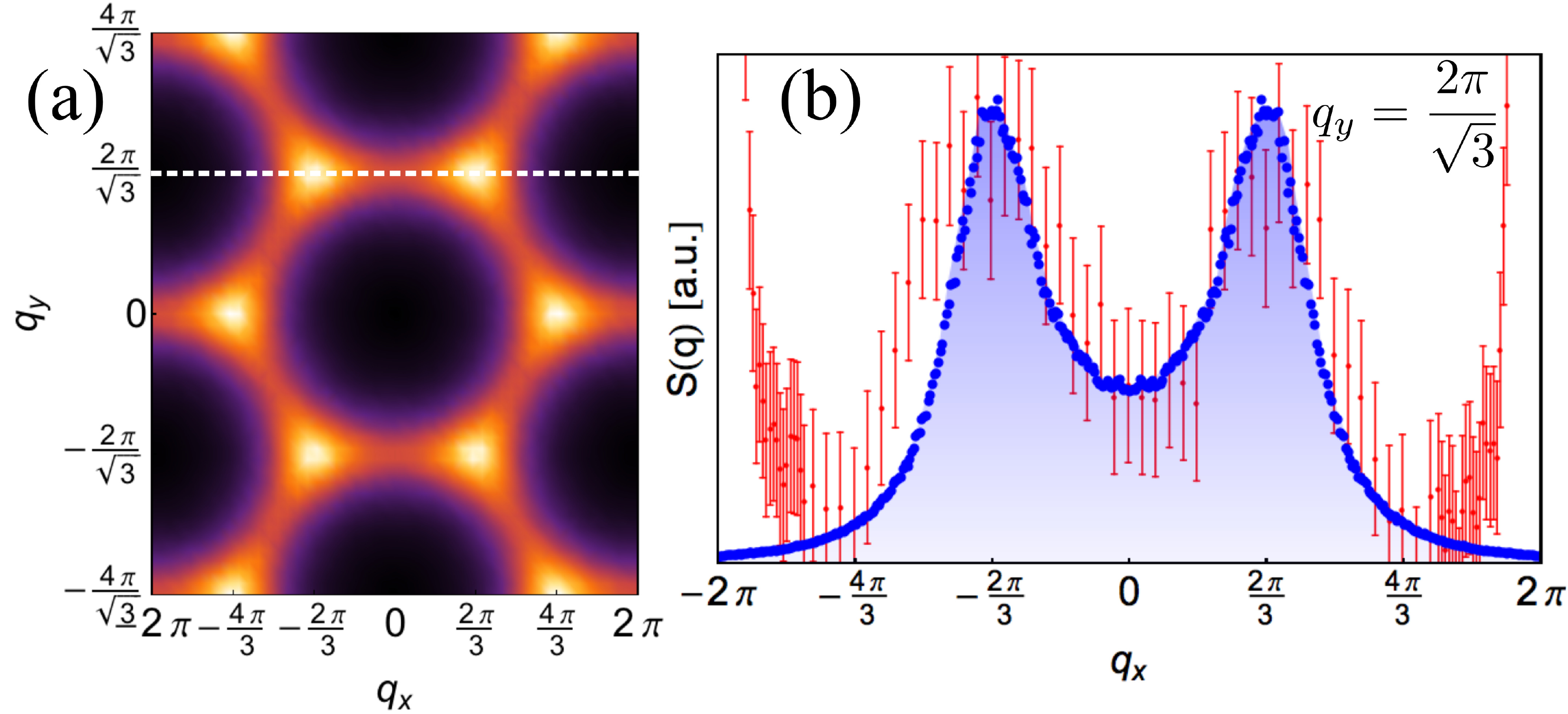}
\caption{\footnotesize{
Comparison between the simulated structure factor following from $E_{\sf Coul} $ [Eq.~\ref{eq:CoulEn}] and x-ray diffraction experiments for Ba$_3$CuSb$_2$O$_9$\cite{nakatsuji12}.
The simulation temperature, $T_{\sf frz}$, is chosen so as to give the best fit to the experimental data and $L=48$.
(a) Simulated structure factor at $T=T_{\sf frz}$ with $q_{\sf z}=\pi$.
(b) Cut through the simulated structure factor at $q_y=\frac{2\pi}{\sqrt{3}}$ and $q_{\sf z}=\pi$ (blue dots, shown by white dashed line in (a)) compared to x-ray diffraction experiments (red dots).
Bragg peaks at $q_x = \pm 2\pi$ are ignored in the simulation, since these are independent of the dumbbell ordering.
}}
\label{fig:strucfac}
\end{figure}

For $T>T_{\sf c}$ we perform simulations to measure the dumbbell structure factor.
In the absence of interaction between bilayers, this is given by,
\begin{align}
S({\bf q}_\perp,q_{\sf z}) = 
\sin^2 \frac{q_{\sf z}}{2}\left| 
\sum_i \sigma_i \mathrm{exp}[{i{\bf q}_\perp \cdot {\bf r}_{\perp,i}}] 
\right|^2,
\label{eq:Sq}
\end{align}
where ${\bf r}_{\perp,i}$ measures the position of dumbbell $i$ in the plane of the triangular lattice.
Here $q_{\sf z} = 2\pi z l /c$, where for Ba$_3$CuSb$_2$O$_9$ the dumbbell height is $z=2.69\AA$, the  unit cell has a height $c=14.37\AA$  and $l$ is measured relative to the structural Bragg peaks\cite{nakatsuji12}.
For $l=0$ it is not possible to observe scattering from the dumbbell structure, as there is a destructive interference between X and Sb ions within the same dumbbell. 
Scattering is strongest when $q_{\sf z}=(2n+1)\pi$, where $n$ is an integer, and for $n=0$ this corresponds to $l=c/(2z) \approx 3$.
Fig.~\ref{fig:ratio} shows $S(q_x,2\pi/\sqrt{3},\pi)$ at a range of temperatures, and there are 
diffuse peaks centred on \mbox{${\bf q}_\perp = (\pm 2\pi/3,2\pi/\sqrt{3})$} [see also Fig.~\ref{fig:strucfac}a].

The diffuse nature of the peaks in $S({\bf q})$ [Eq.~\ref{eq:Sq}] corresponds to the absence of long-range order in the dumbbell structure.
The width of the peaks at half maximum gives a measure of the correlation length, $\xi_{\sf Is}$.
For example, for $T=0.9\psi_{\sf nn}$ [blue curve in Fig.~\ref{fig:ratio}], we find $\xi_{\sf Is} \sim 2a$.
In domains of this lengthscale the system is correlated in a stripe-like pattern [see Fig.~\ref{fig:lattice}].

The motivation for studying the dumbbell structure factor is that it can be compared with low temperature x-ray diffraction data.
This allows the freezing temperature, $T_{\sf frz}$, of the sample to be determined, and then simulation at this temperature can be used to shed light on the low-temperature structure of the dumbbells in the material.
One way to determine $T_{\sf frz}$ is to consider the ratio \mbox{$R=S(0,2\pi/\sqrt{3},q_{\sf z})/S(2\pi/3,2\pi/\sqrt{3},q_{\sf z})$}, since this is sensitive to temperature, as can be seen in Fig.~\ref{fig:ratio}.
The inset to Fig.~\ref{fig:ratio} shows how $R$ increases as a function of $T$, eventually saturating in the  uncorrelated, high-temperature region.

X-ray diffraction data for Ba$_3$CuSb$_2$O$_9$, which is taken from Ref.~[\onlinecite{nakatsuji12}], is shown in Fig.~\ref{fig:strucfac}. 
The value $R\approx 0.4$ is extracted, giving $T_{\sf frz}/\psi_{\sf nn} \approx 0.9$, and the simulated structure factor at this temperature is superposed on the experimental data, showing a good fit.
The freezing temperature can be converted into Kelvin by reintroducing the dimensionful prefactors in $E_{\sf Coul} $ [Eq.~\ref{eq:CoulEn}].
The only unknown is the relative permittivity $\epsilon_r$.
The dumbbells are definitely frozen at $T=300$K, the synthesis temperature is $>1000$K\cite{nakatsuji12}, and, for $T_{\sf frz}$ to be within these limits, a not unreasonable value of $\epsilon_r \sim 10$ is necessary.
Furthermore, the relatively high value of $T_{\sf frz}$ provides a justification for ignoring coupling between bilayers.
However, the fact that diffuse scattering is observed at $l=10$ suggests that some inter-bilayer correlation is present\cite{nakatsuji12}.
This is left for future investigation.


%
\begin{figure}[t]
\centering
\includegraphics[width=0.45\textwidth]{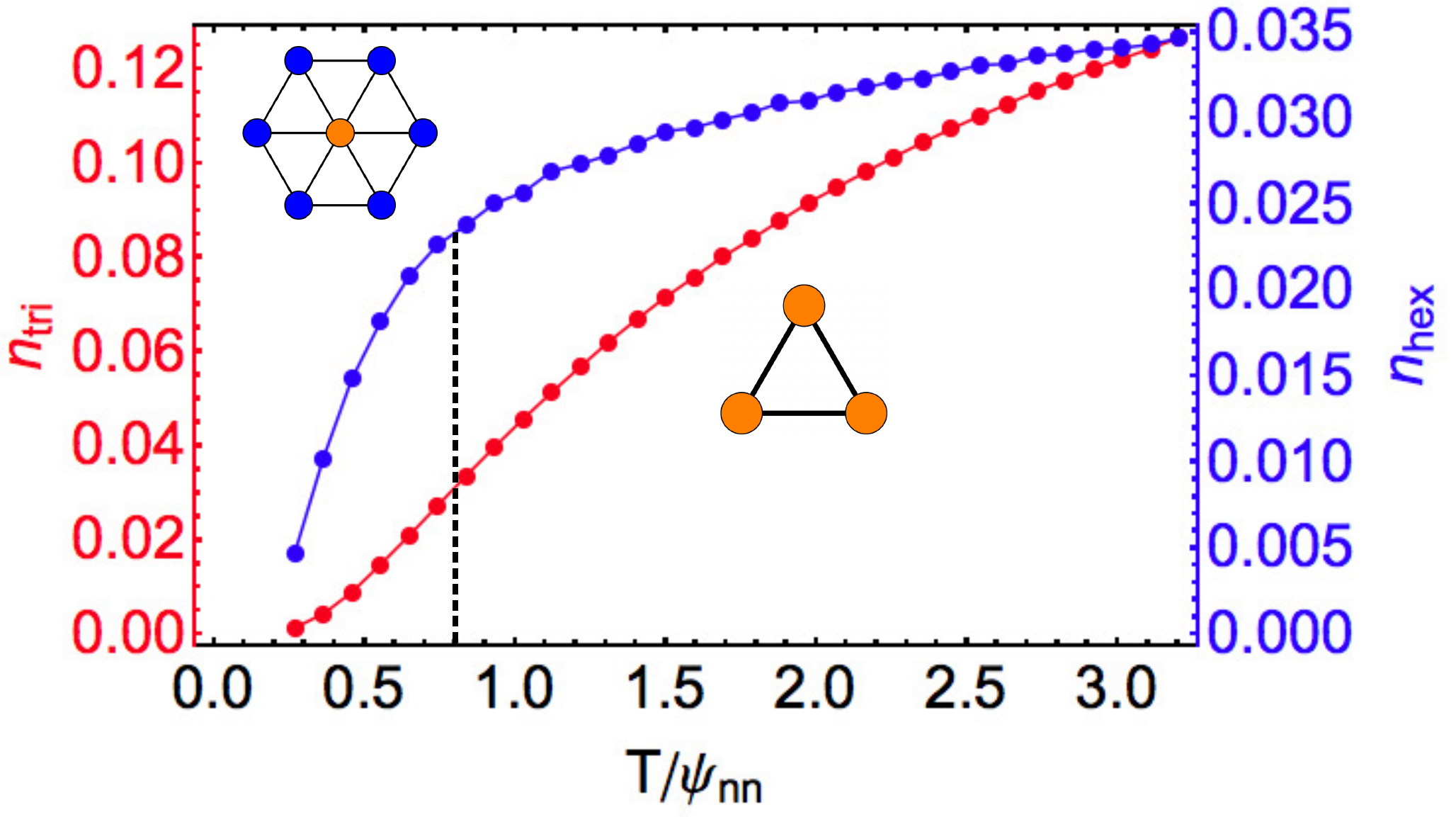}
\caption{\footnotesize{
Fraction of hexagonal plaquettes, $n_{\sf hex}$, and defect triangles, $n_{\sf tri}$, as predicted from simulations of $E_{\sf Coul} $ [Eq.~\ref{eq:CoulEn}].
Hexagonal plaquettes have 6 dumbbells of equivalent orientation surrounding a dumbbell of the opposite orientation. 
The fraction of hexagonal plaquettes relative to a long range honeycomb lattice ($N/3$ plaquettes) rapidly saturates with increasing temperature at $n_{\sf hex} \approx 0.035$ (blue, upper curve).
Defect triangles have three dumbbells with the same orientation, and the fraction relative to a ferromagnetic state ($2N$ defect triangles) steadily increases with temperature (red, lower curve).
The black dashed line shows $T/\psi_{\sf nn}=0.9$, 
which is believed to describe the low-temperature dumbbell structure of the Ba$_3$CuSb$_2$O$_9$ crystals studied in Ref.~[\onlinecite{nakatsuji12}] (see Fig.~\ref{fig:strucfac}).
}}
\label{fig:nhex}
\end{figure}

Once $T_{\sf frz}$ has been determined, the dumbbell structure at this effective lattice temperature can be studied in detail.
The density of defect triangles, $n_{\sf tri}$, on which all three dumbbells are orientated in the same direction, is shown in Fig.~\ref{fig:nhex}.
This density is measured relative to a ferromagnetic state, in which all dumbbells are orientated in the same direction.
The density, $n_{\sf tri}$, increases steadily with temperature and, for $T_{\sf frz}/\psi_{\sf nn}=0.9$, is given by \mbox{$n_{\sf tri} \approx 0.03$}.

Also shown in Fig.~\ref{fig:nhex} is the density of hexagonal plaquettes, $n_{\sf hex}$, measured relative to a long-range honeycomb arrangement of dumbbells ($N/3$ plaquettes).
Hexagonal plaquettes are defined as 6 equivalently orientated dumbbells surrounding a dumbbell of the opposite orientation.
The hexagon plaquette density remains low at all temperatures, rapidly saturating at only $n_{\sf hex} \approx 0.035$, and, for 
$T_{\sf frz}/\psi_{\sf nn}=0.9$, is given by $n_{\sf hex} \approx 0.025$.
In Ref.~[\onlinecite{nakatsuji12}], the presence of diffuse peaks in the x-ray diffraction spectrum at ${\bf q}_\perp=(2\pi/3,2\pi/\sqrt{3})$  [see Fig.~\ref{fig:strucfac}] was taken as proof of a short-range honeycomb arrangement of the dumbbells, since this is the wavevector at which Bragg peaks are found for a long-range ordered honeycomb arrangement.
Here we have shown that such a signal arises even in the absence of a significant number of hexagonal plaquettes, the building blocks of the honeycomb lattice.

How should the lattice of X ions be described, if not by a honeycomb lattice?
To answer this a representative snapshot of the simulations at $T_{\sf frz}/\psi_{\sf nn}=0.9$ is shown in Fig.~\ref{fig:lattice}b.
The lattice can be divided into a set of equally orientated clusters -- that is clusters of neighbouring dumbbells of the same orientation that are completely surrounded by dumbbells of the opposite orientation (shown joined by either blue or orange bonds in Fig.~\ref{fig:lattice}b).
Superexchange between the electronic degrees of freedom associated with the X ions predominantly occurs within these equally orientated clusters, as superexchange between oppositely orientated neighbouring dumbbells is expected to be weak\cite{nakatsuji12}. 
These superexchange-linked clusters can be seen to have a branching structure, and a wide distribution of sizes, $n$.

\begin{figure}[t]
\centering
\includegraphics[width=0.49\textwidth]{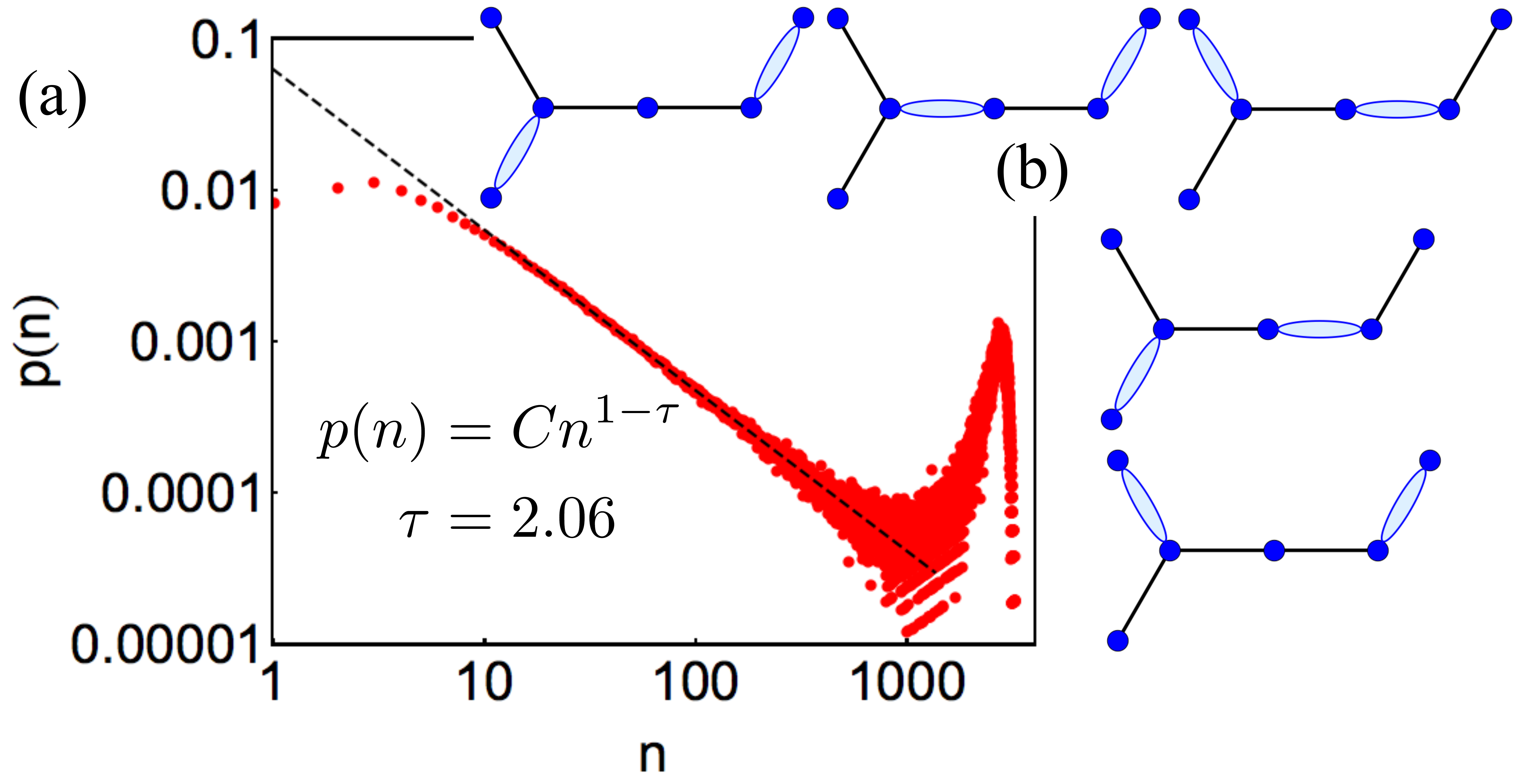}
\caption{\footnotesize{
Statistics of superexchange linked Cu$^{2+}$ clusters at \mbox{$T_{\sf frz}/\psi_{\sf nn}=0.9$}, measured by simulation of $E_{\sf Coul} $ [Eq.~\ref{eq:CoulEn}].
(a) The probability, $p(n)$, that a site belongs to a cluster of size $n$ ($L=48$).
For $10<n<1000$ a power-law distribution $p(n) = 0.063/n^{1.06}$ provides a good fit to the data.
At large cluster sizes ($n > 1000$) the finite size of the simulation becomes important.
(b) A 6-site superexchange linked cluster of Cu$^{2+}$ ions, with 5 distinct maximal dimer coverings.
For geometric reasons, only 4 sites can be covered, leaving 2 monomers (orphan sites).
}}
\label{fig:clusters}
\end{figure}

In Fig.~\ref{fig:clusters} we show $p(n)$, the probability that an arbitrary site is part of an $n$-site cluster.
For $L=48$ (N=6912) and for $10<n<2000$ a good fit to the numerical data is obtained using a power-law probability function, \mbox{$p(n) = Cn^{1-\tau}$}, with $C=0.063$ and $\tau=2.06$.
Finite size effects result in a peak of $p(n)$ at large $n$ and the power law also breaks down at $n \lesssim 10$, where stripe-like correlations between Ising spins suppress the number of small clusters.
Finite size scaling analysis of the average size of the largest cluster shows $\langle  n_{\sf max} \rangle \propto L^{d_{\sf f}}$, where $d_{\sf f}=1.90$ is the fractal dimension [see Appendix~\ref{sec:clusters} for more details].
%

These findings match very well to percolation theory for a model of random site filling on the triangular lattice\cite{stauffer}.
In this model the percolation threshold is at $1/2$-filling, and at this critical point the exponents $\tau=187/91=2.055$ and $d_{\sf f}=91/48=1.896$ are predicted.
The close agreement between these exponents and those found in the simulations suggest that the distribution of sizes of the superexchange linked clusters is at, or very close to, the percolation critical point.
Thus superexchange linked clusters can be expected at all lengthscales [see Appendix~\ref{sec:clusters}].
%


%

It is common in the Ba$_3$XSb$_2$O$_9$ family that a sizeable fraction of the electronic spins are ``orphaned'' and interact only weakly with the rest of the system.
This is observed from a variety of experimental probes and, for X=Cu, 
the percentage of orphan spins has been measured in the range 5-16\% \cite{zhou11,quilliam12,nakatsuji12,do14}.
Neutron scattering studies provide evidence that the Cu spins form nearest-neighbour singlet bonds at low temperature\cite{nakatsuji12}, leading us to consider covering the lattice in singlet dimers. 
Maximally covering the lattice of Cu$^{2+}$ ions with nearest-neighbour singlet dimers leaves a number of orphan spins, due to the geometry of the clusters, and an example of this is shown in Fig.~\ref{fig:clusters}b.
At $T_{\sf frz}/\psi_{\sf nn}=0.9$ the percentage of orphan spins calculated in this way is 6\%, and, for $T>T_{\sf c}$ this is almost independent of both the simulation temperature and the system size [see Appendix~\ref{sec:orphans}].
%

In this dimer picture, clusters with $n=1$ are guaranteed to be an orphan spin, and make up about 15\% of the total orphan spin population.
At low temperature, ESR measures the local environment of the orphan spins\cite{nakatsuji12}, and is therefore biased towards a hexagonal local environment.


It is interesting to speculate about the low temperature spin-orbital state in Ba$_3$CuSb$_2$O$_9$.
Theory suggests that a nearest-neighbour singlet bond is associated with a ferro-orbital alignment between the two sites\cite{nasu13,smerald14,nasu15}.
In order for an orbital resonance to occur, it is therefore necessary for the system to resonate between different singlet coverings of the lattice.
The mechanism for this resonance can arise directly from the superexchange interaction, or from coupling to the lattice\cite{nasu13,smerald14,nasu15}.

For a typical Cu$^{2+}$ superexchange-linked cluster found from solving $E_{\sf Coul} $ [Eq.~\ref{fig:lattice}] at $T=T_{\sf frz}$, there are many possible maximal dimer coverings, which, for geometrical reasons, leave a number of uncovered monomer sites (orphan spins).
An oscillation between different dimer coverings can equivalently be viewed as a hopping of orphan spins around the cluster.
Thus resonance between different dimer configurations of a cluster not only provides a mechanism by which orbitals can resonate, but also suggests that most orphan spins will be delocalised.
%
%
Since the largest superexchange linked cluster diverges in the thermodynamic limit, a resonating state of this type can be designated a spin-orbital liquid on the branch lattice.
In this picture it is the correlated dumbbell disorder that promotes liquid-like behaviour.

To test this picture we performed exact diagonalisation for a spin-orbital Hamiltonian\cite{smerald14} on the 6-site cluster shown in Fig.~\ref{fig:clusters} [see Appendix~\ref{sec:ED} for details].
A trial ground state wavefunction was constructed from the 5 different dimer coverings of the cluster (shown in Fig.~\ref{fig:clusters}), and the overlap with the exact wavefunction was 0.98.
%


It is feasible to check experimentally the premise of this article: that X$^{2+}$-Sb$^{5+}$ dumbbells are flippable at high temperature, freeze as temperature is lowered and that the low-temperature structure can be understood from simulating $E_{\sf Coul} $ [Eq.~\ref{fig:lattice}] at $T_{\sf frz}$.
Since $T_{\sf frz}$ is controlled by the cooling rate, there should be a large difference in the dumbbell structure between a crystal slowly cooled from the synthesis temperature and one that is quenched, and this can be studied by making  x-ray diffraction measurements and extracting the ratio $R$ (see Fig.~\ref{fig:ratio}).
%



In conclusion, we have considered the lattice structure of the Ba$_3$XSb$_2$O$_9$ family, which includes a number of  proposed spin-liquid materials.
By studying a model of charged dumbbells on the triangular lattice using Monte Carlo simulations, we find a non-trivial lattice structure [see Fig.~\ref{fig:lattice}b], in which superexchange linked clusters of X ions form a fractal branching structure.
Focusing in particular on X=Cu, which has been proposed as a spin-orbital liquid, we show that the obtained lattice structure is consistent with x-ray diffraction data.
A simple model of nearest-neighbour singlet covering of the lattice results in a reasonable estimate for the number of orphan spins, and gives rise to a scenario in which correlated dumbbell disorder promotes a spin-orbital liquid state with non-localised orphan spins.


{\it Acknowledgments.}   
We thank Sergey Korshunov for very useful discussions at the beginning of this work.
We also thank Luis Seabra, Ludovic Jaubert and Nic Shannon for advice on Monte Carlo simulation.
We are grateful to Satoru Nakatsuji, Hiroshi Sawa and Naoyuki Katayama for discussions and for sharing their x-ray diffraction data.
We thank the Swiss National Science Foundation and its SINERGIA network ``Mott physics beyond the Heisenberg model'' for financial support.


\appendix


\section{$\psi_{ij}(z)$ interaction matrix}
\label{sec:psimatrix}


Here we study in more detail the form of the interaction matrix $\psi_{ij}(z)$, which appears in Eq.~2 in the main text.
In particular we consider how this changes with $z$, the height of the dumbbells [see Fig.~1a in the main text].
In the main text we use $z=0.46a$, with $a$ the triangular lattice constant, which is taken from the material parameters of Ba$_3$CuSb$_2$O$_9$\cite{nakatsuji12}.

\begin{figure}[h]
\centering
\includegraphics[width=0.49\textwidth]{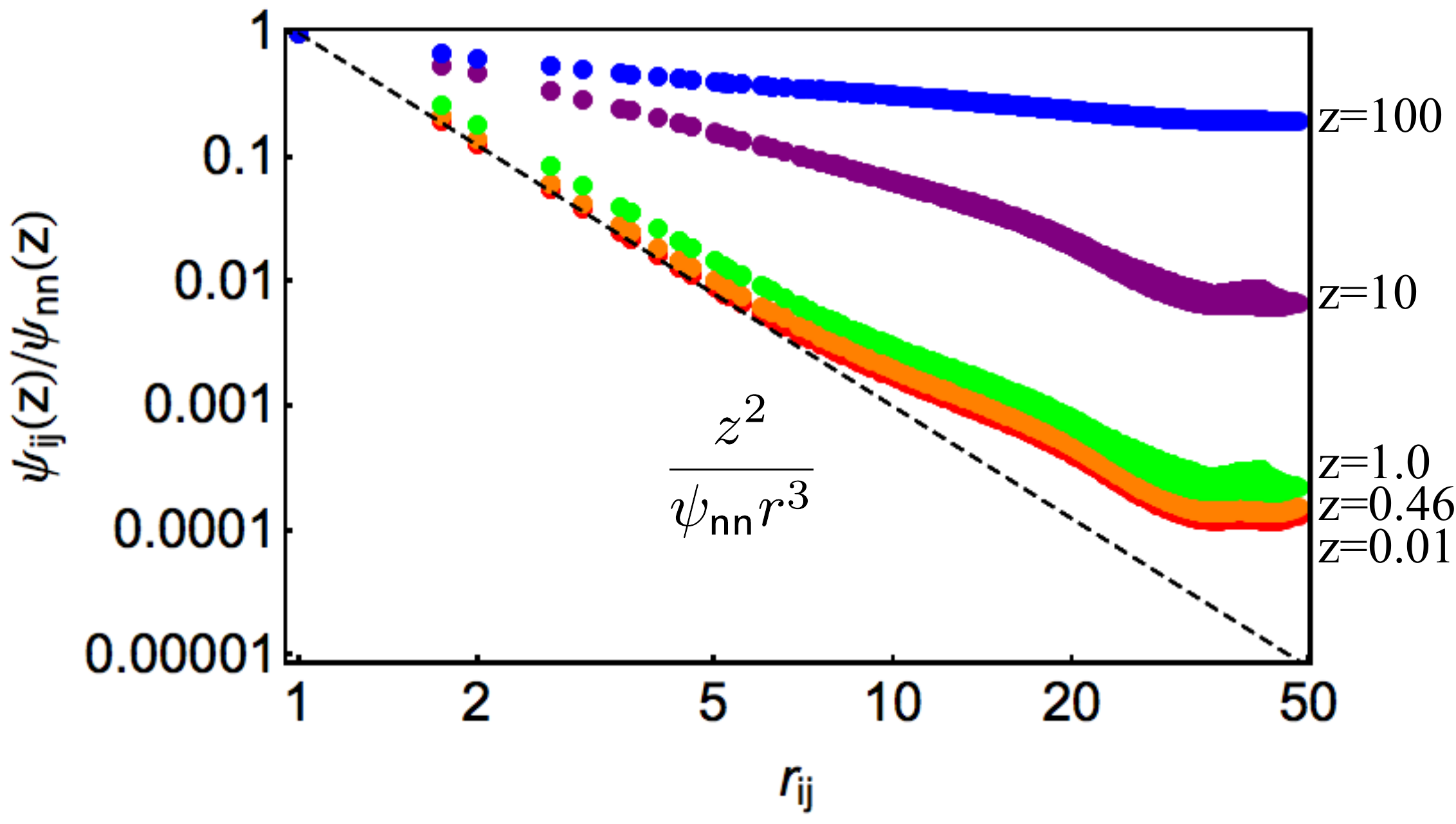}
\caption{\footnotesize{
The interaction matrix $\psi_{ij}(z)/\psi_{\sf nn}(z)$ at various values of $z$ [see Eq.~2 in the main text].
This matrix takes into account the long-range nature of the Coulomb interaction by Ewald summation.
The size of the system is $L=48$ ($N=6912$).
For $z=0.01$ (red), a dipolar interaction is shown as a black dashed line.
Deviations from the dipolar form at large $r_{ij}$ are due to the Ewald summation technique, in which interactions between the central cluster and an infinite set of image clusters are mapped back onto the interactions of the central cluster.
}}
\label{fig:psiijz}
\end{figure}

The energy associated with the interaction of dumbbells on sites $i$ and $j$ is given by,
\begin{align}
E_{ij} = \pm \left( \frac{2}{r_{ij}} - \frac{2}{\sqrt{r_{ij}^2+z^2}}  \right) = \pm \left(\frac{z^2}{r_{ij}^3} +\dots \right),
\end{align}
where the $\pm$ depends on the relative orientation and $r_{ij}$ is the separation.
In the case $r_{ij} \gg z$, the Taylor expansion made in the second equality can be truncated at first order, and the dumbbell-dumbbell interaction has a dipolar form.
If $z \ll a$ the dipolar form will be valid even for nearest-neighbour interactions, while for $z \gg a$, it will only be valid for long-range interactions.

The interaction matrix $\psi_{ij}(z)$ is calculated from $E_{\sf coul}$ [Eq.~1 in the main text] by Ewald summation\cite{grzybowski00}.
The Ewald summation technique allows long-range interactions to be simulated on finite size clusters by tiling the infinite plane with a series of images of the central cluster, and taking into account the interactions with these image clusters.
Therefore $\psi_{ij}(z)$ includes not only the interaction between sites $i$ and $j$ in the original cluster, but also interactions of $i$ in the original cluster with $j$ in all image tiles.

The dependence of $\psi_{ij}(z)$ on the separation $r_{ij}$ is shown in Fig.~\ref{fig:psiijz} for various values of $z$.
Focusing on $z=0.01a$, it can be seen that $\psi_{ij}(z)$ follows a dipolar form at small $r_{ij}$, since the interaction is dominated by the central cluster.
However, at larger $r_{ij}$ the long range interactions become more important with respect to the interactions within the central cluster, and the dipolar form breaks down.


\section{Technical details of Monte Carlo simulations}
\label{sec:MCdetails}


Monte Carlo simulation is used to study $E_{\sf coul}$ [Eq.~2 in the main text].
Here we provide a brief account of the technical details.

A combination of Monte Carlo update methods are employed, including single spin flips, parallel tempering and a worm algorithm based on mapping the nearest-neighbour part of the $\psi_{ij}(z)$ interaction [Eq.~2 in the main text] onto a loop model on the dual honeycomb lattice\cite{zhang09}.
A typical update step includes $9N$ attempts to flip randomly selected individual spins, 10 calls to the worm algorithm and a parallel tempering step.
Clusters are hexagonal in shape and contain $N=3L^2$ lattice sites, where $L$ is the length of the hexagon edges.
Simulations are run using $10^4-10^5$ updates, with equal numbers of thermalisation and measurement steps.
%


\section{Statistics of superexchange linked clusters}
\label{sec:clusters}


Here we consider the superexchange linked clusters of X ions, illustrated in Fig.~1 of the main text and studied in detail in Fig.~5.
The relative orientation of the neighbouring dumbbell degrees of freedom leads to very different superexchange interactions between the X ions.
If two dumbbells are aligned, the superexchange interaction between the associated X ions is comparatively strong, and if they are anti-aligned, the interaction is weak.
This can be seen by comparing the bond angles of the X-O-O-X exchange pathways\cite{nakatsuji12}.
In consequence, one can think of a set of clusters of superexchange linked X ions which, to a first approximation, have no interaction with one another.

Understanding the characteristics of these superexchange linked clusters is important for understanding the physics of the materials.
It is shown in Fig.~4 in the main text that the clusters have very few defect-triangles, in which all three dumbbells are aligned.
Also there are very few hexagonal plaquettes, where 6 aligned dumbbells surround another dumbbell of opposite orientation.
This suggests that the clusters are primarily formed of 1-dimensional chains connected by Y-shape junctions, as can be seen in Fig.~1 in the main text.
An important question to answer concerns the size distribution of these clusters: are there many small clusters or is the system dominated by a few large clusters?
This is equivalent to asking whether or not the clusters percolate.

A useful measure of the cluster size distribution is $p(n)$, the probability that an arbitrary site belongs to a cluster of size $n$.
Since every site must belong to a cluster,
\begin{align}
\sum_{n=1}^N p(n) = 1,
\end{align}
where $N$ is the total number of triangular lattice sites in the system.
One can also define $f_{N}(n)$, the expected frequency of $n$-site clusters in a system of size $N$, as,
\begin{align}
f_{N}(n) = \frac{N p(n)}{n}.
\end{align}
The average cluster size is given by,
\begin{align}
\langle n \rangle = \frac{N}{\langle N_{\sf clus} \rangle} = \frac{N}{\sum_{n=1}^N f_{N}(n)} = \frac{1}{\sum_{n=1}^N p(n)/n} ,
\label{eq:nav}
\end{align}
where $\langle N_{\sf clus} \rangle$ is the average number of superexchange linked clusters.

In order to measure $p(n)$ we use Monte Carlo simulation.
After each Monte Carlo update we split the system into superexchange linked clusters, and determine their size.
Simulations are carried out for system sizes $L=6,12,18,24,36,48$, where $N=3L^2$.
The temperature is set to $T=T_{\sf frz}=0.9 \psi_{\sf nn}$, in order to mimic the crystals synthesised in Ref.~[\onlinecite{nakatsuji12}].

\begin{figure}[t]
\centering
\includegraphics[width=0.49\textwidth]{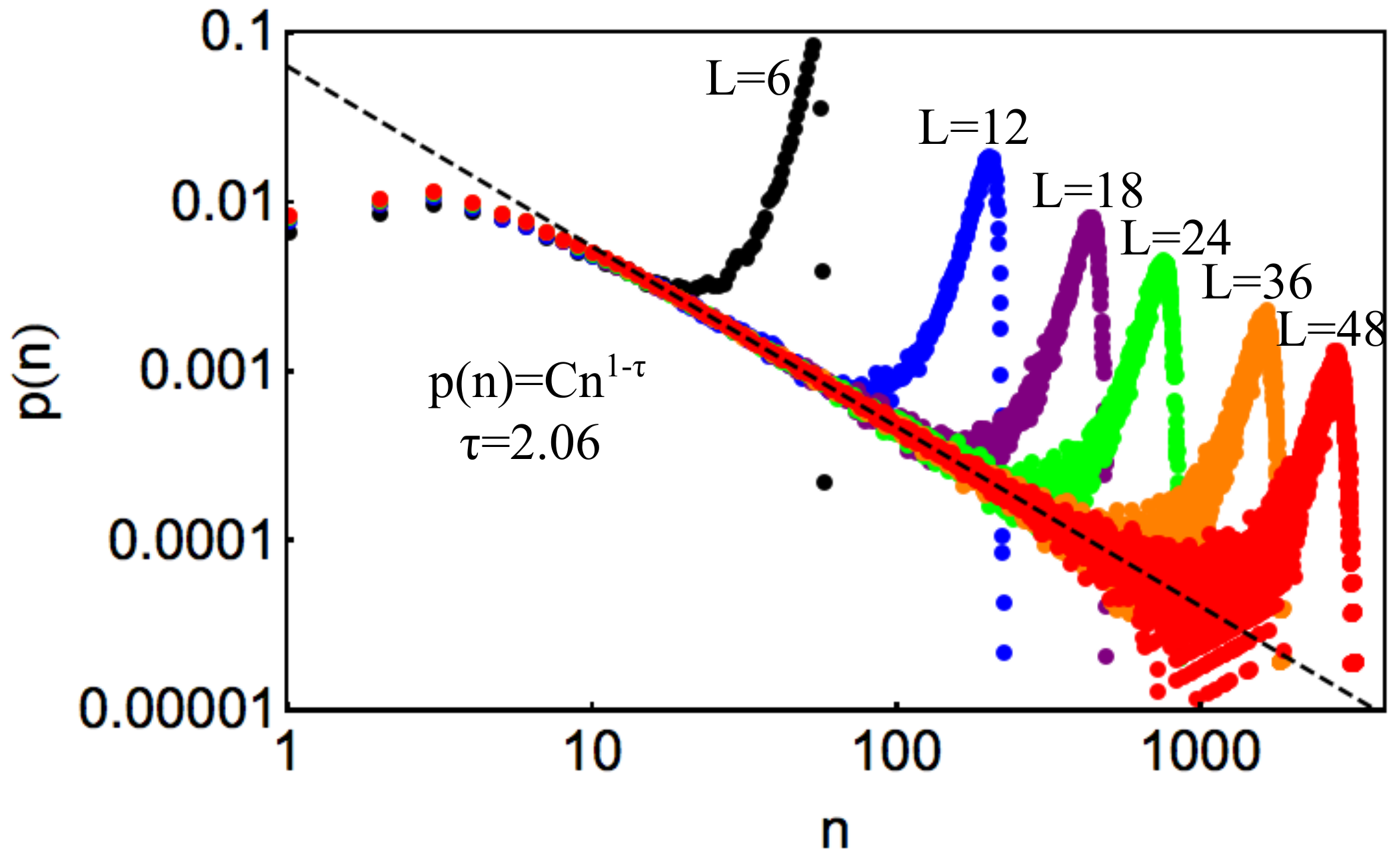}
\caption{\footnotesize{
Statistics of superexchange linked clusters for different system sizes, $L$, measured by Monte Carlo simulation.
The temperature is set to $T=0.9\psi_{\sf nn}$.
The probability that an arbitrary site belongs to an $n$-site cluster is denoted $p(n)$.
For $n \gtrsim 10$ this follows a power law distribution $p(n)=Cn^{1-\tau}$, with $\tau=2.06$, and this is in excellent agreement with percolation theory for a model of random site filling, which predicts $\tau=187/91=2.055\dots$ at the percolation threshold\cite{stauffer}.
The power-law behaviour breaks down at large $n$, due to the finite size of the simulations.
At $n \lesssim 10$ Ising correlations result in a non-power law distribution of cluster sizes.
}}
\label{fig:domstatL}
\end{figure}

The distribution of cluster sizes for different $L$ is shown in Fig.~\ref{fig:domstatL}.
One reason for choosing to measure $p(n)$ is that it is independent of $N$, and simulations at different $L$ collapse onto the same curve.
The finite size of the simulations cuts-off the cluster size at large $n$, and there is a pronounced bump, the position of which is $L$-dependent.

It can be seen from Fig.~\ref{fig:domstatL} that the power law,
\begin{align}
p(n) = C n^{1-\tau},
\end{align}
with $C=0.063$ and $\tau=2.06$, gives a good fit to the Monte Carlo simulations for $n \gtrsim 10$.
The exponent $\tau$ is known as the Fisher exponent\cite{stauffer}.
This power law fit can also be used to demonstrate that the bump at large $n$ is a finite size effect.
As an example, for the $L=48$ ($N=6912$) system the fraction of sites which are part of a cluster with $n>2000$ is given by,
\begin{align}
\sum_{2000}^{6912} p(n) = 0.626,
\end{align}
which compares well to integrating the power law to infinity,
\begin{align}
\int_{2000}^\infty  C n^{1-\tau} = 0.631.
\end{align}
Thus it can be seen that the bump contains all the clusters that in an infinite system would have $n>N/2$, but due to the finite size of the simulation are cut-off at $n<N/2$.

\begin{figure*}[ht]
\centering
\includegraphics[width=0.9\textwidth]{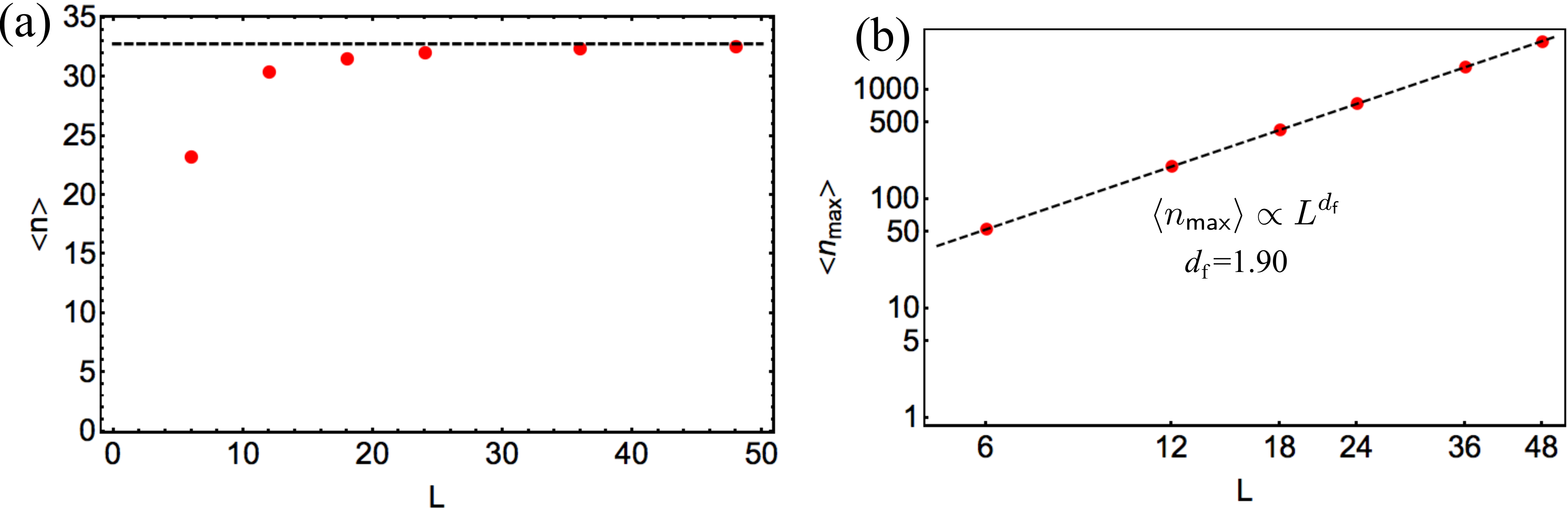}
\caption{\footnotesize{
Average and maximum sizes of superexchange linked clusters, as measured by Monte Carlo simulation.
(a) As the system size $L$ is increased, the average cluster size saturates at $\langle n \rangle = 32.8$.
(b) The average size of the largest cluster grows as $\langle n_{\sf max} \rangle \propto L^{d_{\sf f}}$, with the fractal dimension $d_{\sf f} = 1.90$.
This is in good agreement with percolation theory for random site filling, which predicts $d_{\sf f} = 91/48 = 1.896\dots$ at the percolation threshold\cite{stauffer}.
This suggests that the dumbbell model studied here is tuned to the percolation critical point.
}}
\label{fig:nmax}
\end{figure*}

The average cluster size, $\langle n \rangle$, can be calculated from Eq.~\ref{eq:nav}, and it can be seen in Fig.~\ref{fig:nmax} that it saturates with increasing $L$ at $\langle n \rangle = 32.8$.
More interesting is to study the average size of the largest cluster, $\langle n_{\sf max} \rangle$, as a function of $L$.
As is shown in Fig.~\ref{fig:nmax}, this follows a power law distribution, $\langle n_{\sf max} \rangle \propto L^{d_{\sf f}}$, where we measure $d_{\sf f} = 1.90$ as the fractal dimension\cite{stauffer}.

The simulation results for $\tau$ and $d_{\sf f}$ can be compared to the findings of percolation theory\cite{stauffer}.
For uncorrelated filling of sites in $d=2$, theory predicts that at the percolation threshold $\tau=187/91=2.055\dots$ and $d_{\sf f} = 91/48 = 1.896\dots$ in excellent agreement with the simulations.
For random site filling on the triangular lattice, the percolation threshold is at a filling fraction of $1/2$.

The agreement between the random site filling percolation model and simulations of the Ising system shows that correlation between Ising spins is unimportant at large distances.
This is not surprising, since the temperature is considerably above the Ising critical temperature, at which there is a transition to a low-temperature stripe ordered state.
Since the power law behaviour of $p(n)$ breaks down at $n\lesssim 10$, this gives a rough measure of the Ising correlation length, $\xi_{\sf Is} \sim \sqrt{10}$.
This is a similar to the correlation length $\xi_{\sf Is} \sim 2$ found from fitting a Lorentzian function with full width at half maximum  of $1/\xi_{\sf Is}$ to the Ising structure factor shown in Fig.~2 in the main text.

The power law behaviour of $p(n)$ and $\langle n_{\sf max} \rangle$ holds over all the simulated system sizes.
A minimal conclusion is therefore that the percolation correlation length, $\xi_{\sf p}$, is large compared to the characteristic lengthscale of the largest system size (L=48).
Finite size simulations cannot prove whether the system is exactly tuned to the percolation critical point, where $\xi_{\sf p} \to \infty$.
What one can say is that the ratio of the two dumbbell populations in the Ising model is peaked at 1:1, and this peak becomes sharper as $L$ is increased.
This gives an effective filling fraction of $1/2$, the critical filling fraction of the random site filling percolation model on the triangular lattice\cite{stauffer}.
Thus it appears that the lattice of superexchange linked clusters is very naturally tuned to the percolation critical point.
This result is valid for a large range of temperatures, only breaking down at low-temperature close to the Ising critical point, where $\xi_{\sf Is}$ increases, and Ising correlation effects become important at large lengthscales.
For temperatures above this, $\xi_{\sf Is}$ determines the lengthscale at which the power law behaviour in $p(n)$ breaks down.


\section{Orphan spins}
\label{sec:orphans}


Experimentally, it has been shown that Ba$_3$CuSb$_2$O$_9$ has an orphan spin population of 5-16\% \cite{zhou11,quilliam12,nakatsuji12,do14}.
Here we make a theoretical estimate of the number of orphan spins that follows from the branch lattice structure, illustrated in Fig.~1 of the main text, and find that this is in good agreement with the experimental measurements.

The branch lattice can be split into superexchange linked clusters of dumbbells with the same orientation, as described in detail in Section~\ref{sec:clusters}.
To a first approximation, superexchange interactions between the X ions are strong within the cluster, while the interaction between ions belonging to different clusters can be neglected.
The simple model we propose is that of dimer covering of the clusters.
In this model dimers correspond to nearest-neighbour spin-singlet bonds, with associated ferro-orbital alignment.
Due to the geometry of the clusters, it is in general not possible to cover all sites with dimers, and there will remain some monomer sites [see Fig.~5 in the main text].
These monomers can be interpreted as orphan spins.
The idea of nearest-neighbour spin-singlet covering of the clusters arises from both experiment\cite{nakatsuji12,quilliam12} and theory\cite{nasu13,smerald14}. 

\begin{figure*}[ht]
\centering
\includegraphics[width=0.9\textwidth]{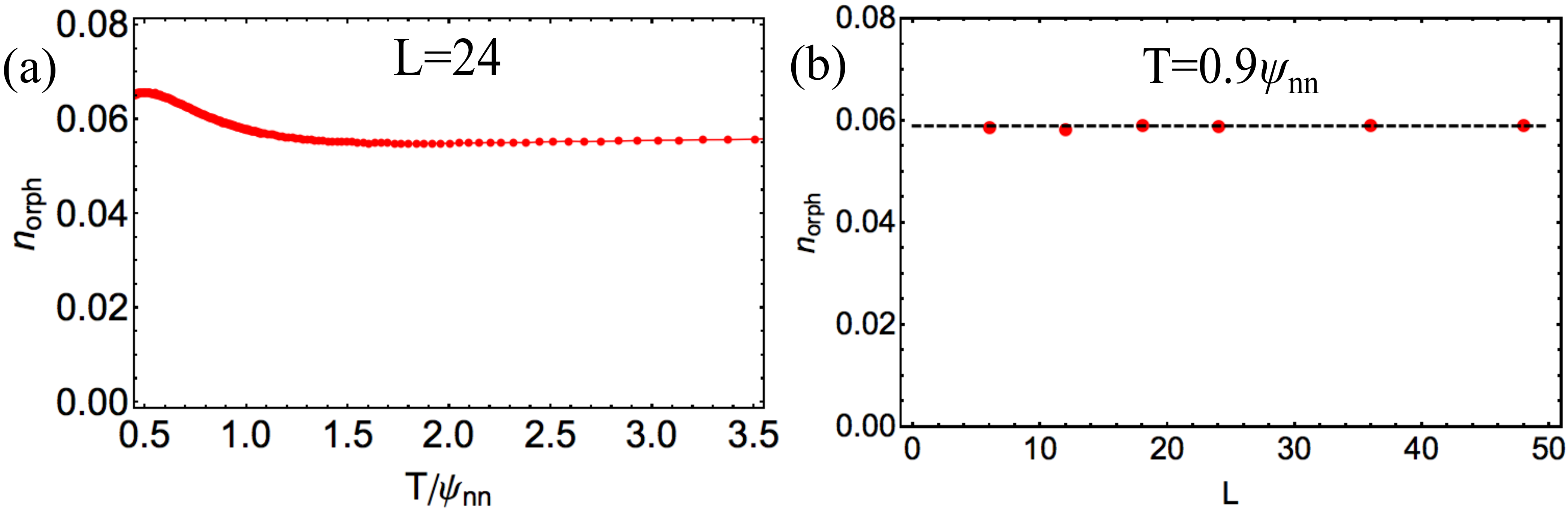}
\caption{\footnotesize{
Fraction of orphan spins, $n_{\sf orph}$, as measured by Monte Carlo simulation.
Superexchange linked clusters are maximally covered in dimers, representing nearest-neighbour spin singlets, leaving a number of monomer spins that are considered to be orphans.
(a) For constant system size, $L=24$, the fraction of orphan spins is only weakly temperature dependent.
(b) At constant temperature, $T=0.9\psi_{\sf nn}$, the fraction of orphan spins is given by $n_{\sf orph}=0.059$, independent of the system size, $L$.
}}
\label{fig:norph}
\end{figure*}

The problem we thus wish to solve is how to maximally cover an irregular cluster of sites selected from the triangular lattice with dimers.
First we perform Monte Carlo simulations of the Ising model given in Eq.~2 of the main text.
After each Monte Carlo update the lattice is split into superexchange linked clusters, and we find a power law distribution of cluster sizes, as shown in Fig.~\ref{sec:clusters}.
For each of these clusters we use a numerical algorithm to maximally cover it in dimers, and determine the number of monomer sites remaining.

The algorithm is as follows. 
First each site of the lattice is assigned a connectivity, which is just the number of nearest-neighbour sites on the triangular lattice that are part of the same cluster.
This connectivity is in the range $1-6$, though in practice connectivities of $4-6$ are relatively uncommon.
We select at random one of the sites with the lowest connectivity.
If this has connectivity 1 (the usual case) we place a dimer between the chosen site and its neighbouring site, delete both sites from the cluster and recalculate the connectivitites of the remaining sites.
If the smallest connectivity is 2 or greater, a dimer is placed on the bond connecting the selected site and a randomly chosen second site.
These sites are then removed from the cluster and the connectivities recalculated.
After the recalculation, it sometimes happens that there will be sites with connectivity 0, and these are assigned to be monomers.
The same process is repeated until all sites have been covered by a dimer or assigned as a monomer.

We do not have a mathematical proof of this algorithm.
For small cluster sizes we have tested it against the slow but foolproof method of enumerating all dimer coverings and finding which one(s) cover the most sites.
We have also constructed a number of apparently pathological cases, and determined that the algorithm correctly calculates the maximal dimer covering.

The fraction of orphan spins, $n_{\sf orph}$, calculated by this method is shown in Fig.~\ref{fig:norph} as a function of both temperature, $T$, and of system size $L$.
For fixed system size, $L=24$, it can be seen that $n_{\sf orph}$ varies very little with increasing temperature.
For fixed temperature, $T=0.9\psi_{\sf nn}$, it can be seen that $n_{\sf orph} = 0.059$ is essentially independent of $L$.
This value is in good agreement with experimental measurements in Ba$_3$CuSb$_2$O$_9$ of $n_{\sf orph} = 0.05-0.16$ \cite{zhou11,quilliam12,nakatsuji12,do14}.


\section{Spin-orbital state of a 6-site cluster}
\label{sec:ED}


In order to test the nature of the spin and orbital state of superexchange linked clusters of Cu in Ba$_3$CuSb$_2$O$_9$ we perform exact diagonalisation of a spin-orbital Hamiltonian on a 6-site cluster.

The Hamiltonian we consider was derived in Ref.~[\onlinecite{smerald14}] from a 2-orbital Hubbard model, and is given by,
\begin{align}
&\mathcal{H}_{\sf ST} =  \sum_{\langle ij \rangle} 
\left\{
-\mathcal{P}_{ij}^{S=0}
\left[ -\frac{2}{3} {\bf T}_i\cdot {\bf T}_j  
 +\frac{16}{9} ({\bf n}_{ij}\cdot {\bf T}_i)({\bf n}_{ij} \cdot {\bf T}_j) \right. \right. \nonumber \\
& \qquad \qquad \left. -\frac{8}{9} ({\bf n}_{ij}\cdot {\bf T}_i + {\bf n}_{ij} \cdot {\bf T}_j)
+\frac{5}{6}
\right]   \nonumber \\
& +\left.
\mathcal{P}_{ij}^{S=1}
\left[ -\frac{2}{3} {\bf T}_i\cdot {\bf T}_j 
 +\frac{16}{9} ({\bf n}_{ij}\cdot {\bf T}_i)({\bf n}_{ij} \cdot {\bf T}_j) 
- \frac{5}{18}
\right]
\right\},
\label{eq:HST-honeycomb}
\end{align}
where,
\begin{align}
\mathcal{P}_{ij}^{S=0} = \frac{1}{4} -{\bf S}_i\cdot {\bf S}_j , \quad
\mathcal{P}_{ij}^{S=1} = \frac{3}{4} + {\bf S}_i\cdot {\bf S}_j ,
\end{align}
are singlet and triplet projection operators and ${\bf T}_i$ is an orbital pseudospin-1/2, with $T^{\sf z}=1/2$ representing a $d^{3{\sf z}^2 -{\sf r}^2}$ orbital and $T^{\sf z}=-1/2$ a $d^{{\sf x}^2 -{\sf y}^2}$ orbital.
The vector ${\bf n}_{ij}$ is different for the 3 possible bond orientations (see Fig.~1 in the main text for bond labelling) and given by,
\begin{align}
{\bf n}_{ij\in {\sf A}} &= (0,0,1)  \nonumber \\
{\bf n}_{ij\in {\sf B}} &= \left( \frac{\sqrt{3}}{2},0,-\frac{1}{2} \right)  \nonumber \\
{\bf n}_{ij\in {\sf C}} &=  \left( -\frac{\sqrt{3}}{2},0,-\frac{1}{2} \right).
\label{eq:nvectors}
\end{align}
$\mathcal{H}_{\sf ST}$ [Eq.~\ref{eq:HST-honeycomb}] is equivalent to Eq.~8 in Ref.~[\onlinecite{smerald14}] with the Hund's rule coupling set to $J=0$ and the ratio of hopping parameters set to $t/t^\prime = -1/3$, which is believed to be relevant for Ba$_3$CuSb$_2$O$_9$\cite{smerald14}.

\begin{figure}[ht]
\centering
\includegraphics[width=0.49\textwidth]{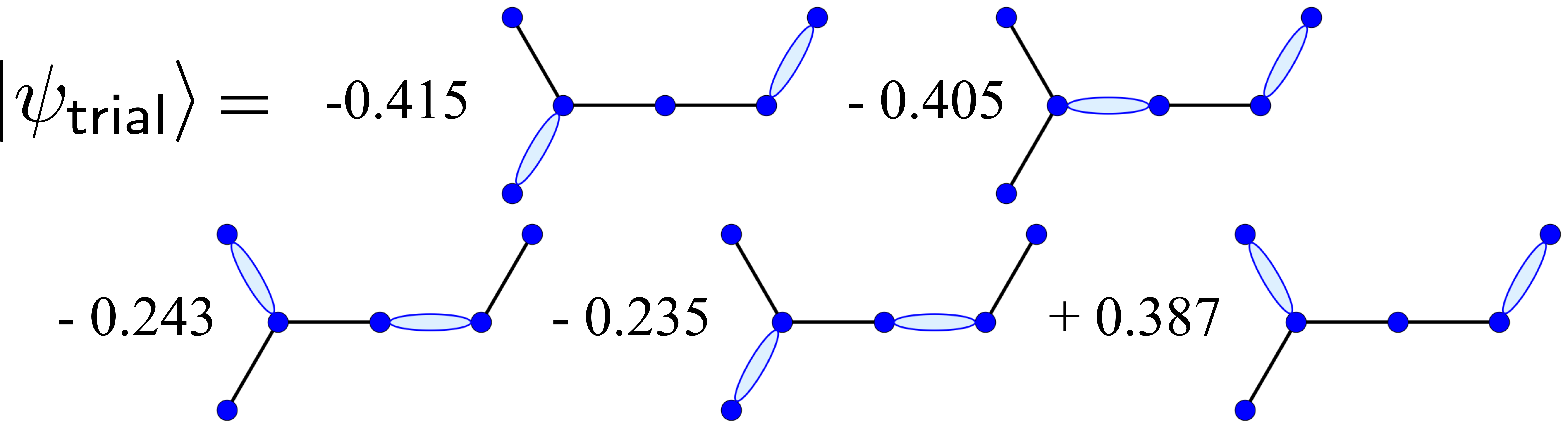}
\caption{\footnotesize{
Trial spin-orbital wavefunction, $|\psi_{\sf trial} \rangle$, for $\mathcal{H}_{\sf ST}$ [Eq.~\ref{eq:HST-honeycomb}] on a 6-site cluster.
Blue ellipses represent spin singlet bonds with ferro-orbital alignment in the $d^{{\sf x}^2 -{\sf y}^2}$ orbitals (and $2\pi/3$ rotations).
Unpaired sites can be thought of as orphan spins, and are found to have a $Q_2$ type Jahn-Teller distortion (see Fig.~\ref{fig:octdistortions}).
The overlap with the exact wavefunction, $| \psi_{\sf ED} \rangle$, is found to be $\langle \psi_{\sf trial} | \psi_{\sf ED} \rangle = 0.98$, showing that the trial wavefunction gives a good description of the exact ground state.
}}
\label{fig:EDwfn}
\end{figure}

We perform exact diagonalisation of $\mathcal{H}_{\sf ST}$ [Eq.~\ref{eq:HST-honeycomb}] on the 6-site cluster shown in Fig.~\ref{fig:EDwfn}.
The idea of choosing such a cluster is that it cannot be fully covered with dimers, and therefore one can explore the nature of the orphan spins (see Section~\ref{sec:orphans}).
\begin{figure}[h]
\centering
\includegraphics[width=0.4\textwidth]{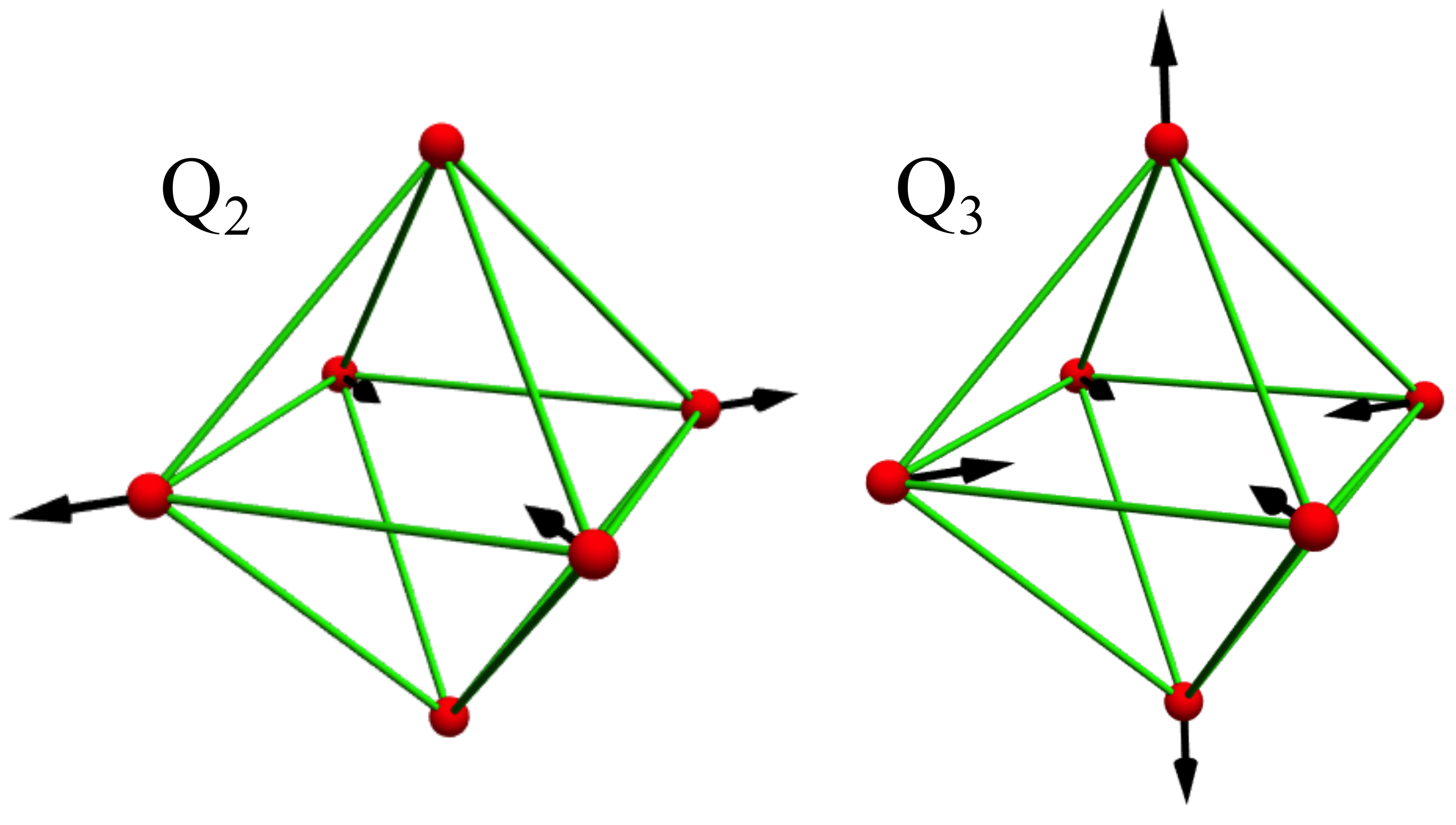}
\caption{\footnotesize{
Jahn-Teller distortions of the oxygen octahedra associated with $e_{\sf g}$ orbitals.
A $Q_3$ type distortion is expected when the orbital state is $T^{\sf z}=\pm 1/2$ (and $2\pi/3$ rotations).
A $Q_2$ type distortion is expected when the orbital is in the $T^{\sf x}=\pm 1/2$ state (and $2\pi/3$ rotations).
}}
\label{fig:octdistortions}
\end{figure}

The exact ground state wavefunction, $| \psi_{\sf ED} \rangle$, is found to be in the $S=0$ sector.
In order to elucidate the physical nature of this exact wavefunction, we construct a trial wavefunction, $|\psi_{\sf trial} \rangle$, shown in Fig.~\ref{fig:EDwfn}.
In the trial wavefunction blue ellipses represent nearest-neighbour spin singlets with ferro-orbital alignment of the two orbitals in either $d^{{\sf x}^2 -{\sf y}^2}$ (A-bonds), $d^{{\sf y}^2 -{\sf z}^2}$ (B-bonds) or $d^{{\sf z}^2 -{\sf x}^2}$ (C-bonds).
Unpaired sites contain an orphan spin, and it is found that the orbital degree of freedom on these sites favours a $Q_2$ type distortion ($T^{\sf x}=\pm 1/2$ and $2\pi/3$ rotations), as opposed to the $Q_3$ type distortion ($T^{\sf z}=\pm 1/2$ and $2\pi/3$ rotations) favoured by orbitals associated with singlet bonds (see Fig.~\ref{fig:octdistortions}). 
We find $\langle \psi_{\sf trial} | \psi_{\sf ED} \rangle = 0.98$, showing that the trial wavefunction gives a good description of the exact ground state.

These findings support the picture of the spin-orbital state given in the main text. 
That is, resonating valence bonds of nearest-neighbour, ferro-orbital spin singlets coexisting with mobile, weakly-coupled orphan spins.
The finding that orphan spins are associated with a $Q_2$ type of Jahn-Teller distortion could be important for understanding the diffuse x-ray scattering experiments reported in Ref.~[\onlinecite{ishiguro13}].

\bibliographystyle{apsrev4-1}
\bibliography{bibfile}


\end{document}